# Structural and Electrical Properties of MoTe$_2$ and MoSe$_2$ Grown by Molecular Beam Epitaxy

Anupam Roy[1*], Hema C. P. Movva[1*], Biswarup Satpati[2], Kyounghwan Kim[1], Rik Dey[1], Amritesh Rai[1], Tanmoy Pramanik[1], Samaresh Guchhait[1#], Emanuel Tutuc[1] and Sanjay K. Banerjee[1]

[1]Microelectronics Research Center, The University of Texas at Austin, Austin, Texas 78758, USA

[2]Surface Physics and Material Science Division, Saha Institute of Nuclear Physics, 1/AF, Bidhannagar, Kolkata 700 064, India

*Address correspondence to anupam@austin.utexas.edu; hemacp@utexas.edu

#Presently at: Department of Physics, University of Maryland, College Park, MD 20742, USA

ABSTRACT: We demonstrate the growth of thin films of molybdenum ditelluride and molybdenum diselenide on sapphire substrates by molecular beam epitaxy. *In situ* structural and chemical analyses reveal stoichiometric layered film growth with atomically smooth surface morphologies. Film growth along the (001) direction is confirmed by X-ray diffraction, and the crystalline nature of growth in the 2H phase is evident from Raman spectroscopy. Transmission electron microscopy is used to confirm the layered film structure and hexagonal arrangement of surface atoms. Temperature dependent electrical measurements show an insulating behavior which agrees well with a two-dimensional variable-range hopping model, suggesting that transport in these films is dominated by localized charge-carrier states.

## INTRODUCTION

Semiconducting two-dimensional (2D) transition metal dichalcogenides (TMDs) are being actively investigated as the channel material for post-Si electronic devices.[1–3] Their atomically thin nature, coupled with unique physical[4,5] and optoelectronic properties,[6–8] have also enabled exploration of novel spin and valleytronic phenomena.[9–11] To date, most of the studies on TMDs have utilized micrometer sized flakes that are mechanically exfoliated from bulk crystals.[12–17] While providing a workable platform for proof-of-concept demonstrations, this approach is un-scalable and prevents large scale integration that is crucial for practical TMD-based device architectures. Chemical vapor deposition (CVD) has recently attracted significant attention for pseudo large area growth of TMD thin films.[18–20] However, the discontinuous nature of the films, lack of thickness control and homogeneity are its major drawbacks. Moreover, CVD growth has been mostly limited to the light chalcogen based TMDs (sulfides and selenides).

Molecular beam epitaxy (MBE) is a versatile growth technique which is widely used for large area growth of high quality crystalline films and heterostructures. Compared to CVD, MBE offers greater control of film thickness, the ability to incorporate dopants, and the capability to grow high quality heterostructures with abrupt interfaces. In the context of TMDs, weak van der Waal's bonding, and the absence of dangling bonds along the *z*-direction enables their crystalline growth on arbitrary substrates through van der Waal's epitaxy (vdWE). Recent reports on vdWE of molybdenum diselenide (MoSe$_2$) permitted investigation of its structural and physical properties using transmission electron microscopy (TEM), photoelectron spectroscopy, and scanning tunneling microscopy.[7,8,21–25] The growth on conducting graphite substrates has however prevented electrical characterization of these films. Furthermore, heavy chalcogen based TMDs such as molybdenum ditelluride ($MoTe_2$) with a bandgap closer to Si (~ 1.0 eV) have received lesser attention, with the MBE growth of $MoTe_2$ being virtually unexplored.

In this work, we report on the structural and electrical properties of large area $MoTe_2$ and $MoSe_2$ thin films grown by MBE on $c$-$Al_2O_3$(0001) (sapphire) substrates. Several *in situ* and *ex situ* characterization techniques are used to confirm the stoichiometry and crystalline nature of the films. Growth on insulating sapphire substrates enables electrical characterization of the as-grown films. Temperature-dependent transport measurements display a trend of increasing resistance with decreasing temperature. The behavior fits well with a 2D Mott variable-range hopping (VRH) mechanism, indicating transport that is dominated by localized charge-carrier states, possibly due to disorder and defects.[26]

**RESULTS AND DISCUSSION**

We chose insulating crystalline sapphire substrates due to their hexagonal surface symmetry, and to enable transport measurements of the as-grown films. The growths were done in a custom-built MBE chamber (Omicron systems) with a base pressure of $1 \times 10^{-10}$ mbar. Single crystal $c$-$Al_2O_3$(0001) substrates were prepared by resistive heating, and monitored by *in situ* reflection high-energy electron diffraction (RHEED). Thin films of $MoTe_2$ and $MoSe_2$ were subsequently grown by co-evaporation of Mo, Te and Mo, Se, respectively. The substrate temperature was 350 °C for $MoTe_2$ and 250 °C for $MoSe_2$. More details of the growth process are described in the Materials and Methods section. Figure 1 shows the RHEED images of the substrates before and after growth. Optical micrographs of the large area, homogeneous (~ 6 × 5 $mm^2$) films of $MoTe_2$ and $MoSe_2$ are shown in Supporting Information S1. The RHEED pattern of the sapphire substrate disappears completely within a few minutes of growth, and a diffused pattern appears, that is indicative of a poly-crystalline film structure (S2 in Supporting Information). Following growth, a high temperature anneal was done at 600 °C for 10 min to improve the crystallinity of the films, resulting in the final sharp, streaky RHEED patterns as shown in Figure 1(b) and (d). The post-growth anneals were performed in a chalcogen rich environment to minimize vacancies due to chalcogen out-diffusion. The $MoTe_2$ RHEED images show a faint signature of spots along with streaks, possibly due to regions of clustered growth and/or disordered surface regions.[21] The streaks in the $MoSe_2$ RHEED images are relatively sharper. We elaborate on this difference in the later sections. Despite the large lattice mismatch with the substrate, the streaky RHEED features suggest well-structured film growth with high crystalline quality and atomically flat surfaces. We also note that the RHEED features are insensitive to both sample and beam orientation.[7] Furthermore, the growth occurs along the $c$-axis, as expected for the vdWE growth of a hexagonal thin film on an hcp(0001) substrate. We focus on data from few-layer films in the subsequent sections in order to ensure film continuity and enable electrical characterization.

X-ray diffraction (XRD) was done to evaluate the structure of the films and to confirm their epitaxial nature. Figure 1(e) and (f) show the XRD patterns for a 4 nm film of $MoTe_2$ and a 5 nm film of $MoSe_2$, respectively. Both the patterns show characteristic peaks that correspond to diffraction from the family of (002) planes, as expected for the symmetry group $P6_3/mmc$. The absence of peaks other than the (00$l$) family confirms vdWE growth along the $c$-axis of the sapphire substrates. The sharp peak at $2\theta = 41.7°$ corresponds to reflection from the (006) plane of the sapphire substrate. The peak positions match well with the (00$l$) planes of 2H-$MoTe_2$ (JCPDS No. 73-1650) and 2H-$MoSe_2$ (JCPDS No. 77-1715).[27,28] The relatively higher intensity of the (002) peak is a signature of a well-stacked layered structure. It is to be noted that the broadness of the peaks is likely due to the thin-film nature; however, disorder or textured nature of the films cannot be ruled out. The extracted $c$-axis lattice constants are 13.9 Å for $MoTe_2$ and 13.0 Å for $MoSe_2$, which match closely with the bulk crystal values.[27,28]

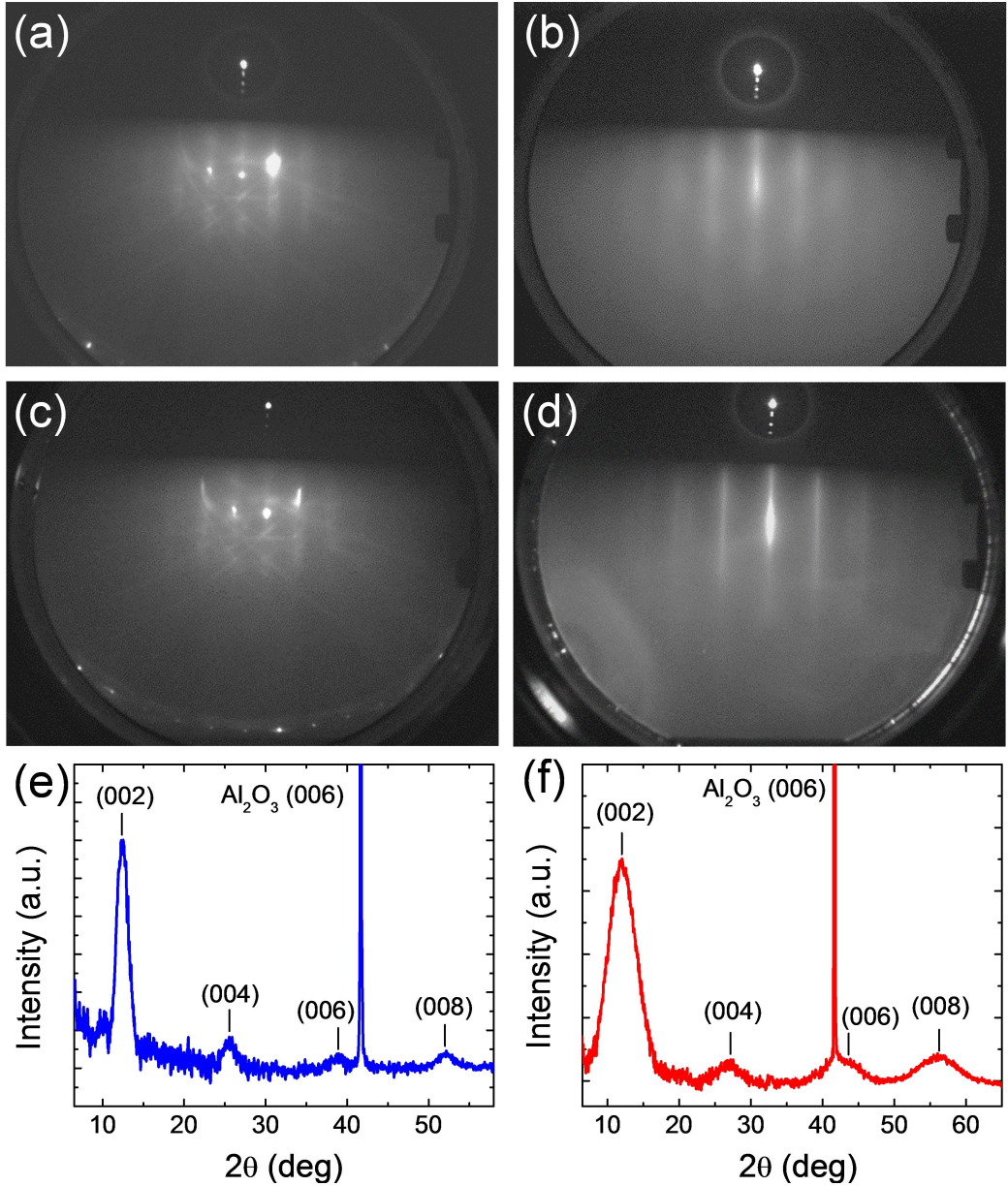

**Figure 1** RHEED patterns (a) before and (b) after growth of MoTe$_2$ on *c*-Al2O3(0001) taken along the [1 1 -2 0] direction show streaky features indicative of layered film growth with atomically flat surfaces. The corresponding patterns for MoSe$_2$ (c, d) show very similar features. The XRD patterns for MoTe$_2$ (e) and MoSe$_2$ (f) show peaks characteristic of the 2H polytypes, and confirm growth along the *c*-axis.

The elemental composition and stoichiometry of the films were investigated through X-ray photoelectron spectroscopy (XPS) measurements. Survey spectra of the films (Supporting Information S3) show predominant peaks from Mo and Te/Se, along with weak contribution from Al and O from the sapphire substrate. Figure 2 shows the high-resolution XPS spectra for both MoTe$_2$ and MoSe$_2$. It is to be noted that the O contribution is only from the substrate, as evident from the absence of oxide doublets in the high-resolution spectra. The binding energies of Mo-3$d_{5/2}$ (227.8 eV) and Mo-3$d_{3/2}$ (230.9 eV) in Figure 2(a) are consistent with the +4 oxidation state of Mo. Similarly, Te-3$d_{5/2}$ (572.4 eV) and Te-3$d_{3/2}$ (582.8 eV) in Figure 2(b) correspond to the -2 oxidation state of Te. There is a close agreement of the peak positions with reported values for 2H-MoTe$_2$ bulk crystals.[29] Absence of Mo and Te peaks of other oxidation states confirms homogeneity of the 2H phase to within the detection limit. Using the integrated peak areas of Mo-3$d$ and Te-3$d$, we obtain a Te:Mo ratio of 1.91. The deviation from the expected stoichiometric value of 2 for 2H-MoTe$_2$ could be due to the presence of disorder possibly from chalcogen vacancies, and/or the presence of other polytypes. The energy difference $\Delta E$ between Mo-3$d_{5/2}$ and Te-3$d_{5/2}$ in MoTe$_2$ is 344.7 eV, which is very close to that of the corresponding elemental peaks (345.2 eV).[29] This is due to the small electronegativity difference between Mo and Te ($\Delta\chi_{Te-Mo}$ = 0.3). As a result, the bonding energy of MoTe$_2$ is lower compared to other Mo based TMDs, and can therefore lead to a higher incidence of Te vacancies. Figure 2(c) and (d) show the corresponding high-resolution XPS spectra of Mo and Se, respectively, in MoSe$_2$. While the Mo spectrum is similar to that of MoTe$_2$, the Se-3$d$ doublet is not as well resolved as the Te-3$d$ doublet. This is due to the lower atomic mass of Se which results in a lower spin splitting of Se-3$d$ compared to Te-3$d$. A Se:Mo ratio of 1.95 is extracted for the MoSe$_2$. The slight deviation from the stoichiometric value of 2 could be due to Se vacancies, similar to Te vacancies in MoTe$_2$. The bonding energy between Mo and Se is higher than Mo and Te, as evident from the $\Delta E$ = 174.4 eV between Mo-3$d_{5/2}$ and Se-3$d_{5/2}$ in MoSe$_2$, compared to 172.4 eV between elemental Mo and Se.[30] A closer to ideal stoichiometry in MoSe$_2$ compared to MoTe$_2$ is due to the stronger Mo-Se bonds. The weaker Mo-Te bonds are also responsible for the higher prevalence of disorder in MoTe$_2$, as apparent in the RHEED patterns of Figure 1.

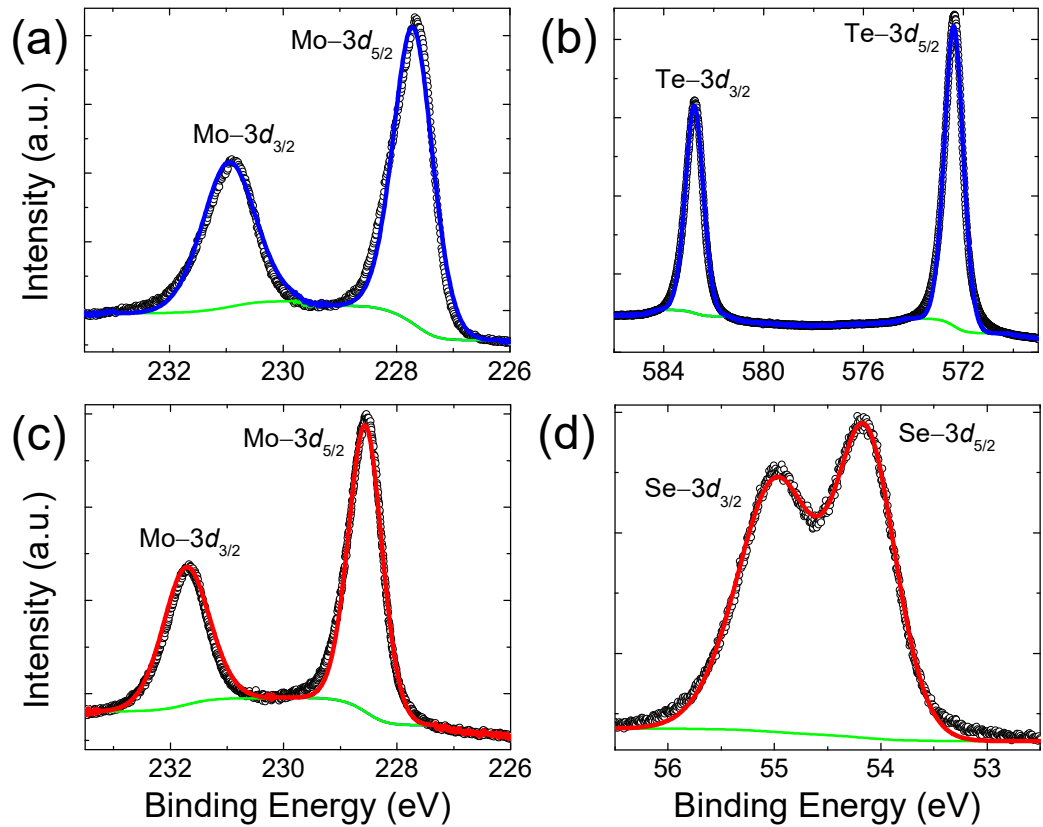

**Figure 2** The high-resolution XPS spectra of MoTe$_2$ and MoSe$_2$ thin films. (a) Mo-3d and (b) Te-3d core level peaks of MoTe$_2$ and (c) Mo-3d and (d) Se-3d core level peaks of MoSe$_2$ show chemical shifts corresponding to homogeneous phases of 2H-MoTe$_2$ and 2H-MoSe$_2$, respectively. An Mo:Te ratio of 1.91 and Mo:Se ratio of 1.95 is extracted from the area fit (solid lines) to the experimental data (circles).

Figure 3(a) and (b) show Raman spectra of the MoTe$_2$ and MoSe$_2$ films, respectively. Spectra from exfoliated bulk flakes on Si/SiO$_2$ substrates are also shown for comparison. The measurements were done using a 532 nm laser. Group theory analysis for bulk TMDs (belonging to the D$_{6h}$ group) predicts four Raman-active and two Raman-inactive modes.[31] Out of the four Raman-active modes, E$_{1g}$, E$^1_{2g}$ and E$^2_{2g}$ arise from the in-plane Raman-active vibrations, and A$_{1g}$ is due to the out-of-plane vibrations. In addition, the Raman-inactive breathing mode B$^1_{2g}$ becomes active in few-layer 2H-TMDs, due to the breaking of crystal symmetry along the c-axis.[32] Both MoTe$_2$ and MoSe$_2$ show prominent E$^1_{2g}$ (MoTe$_2$: 236.1 cm$^{-1}$ and MoSe$_2$: 286.4 cm$^{-1}$) and A$_{1g}$ (MoTe$_2$: 173.9 cm$^{-1}$ and MoSe$_2$: 245.5 cm$^{-1}$) peaks, reaffirming the crystalline layered structure of our films. The positions and relative intensities of both peaks match very well with that of the corresponding exfoliated flakes, as well as previous literature reports.[22,32] The broadening of the peaks in our MBE films compared to exfoliated flakes could be due to defects arising from chalcogen vacancies and/or grain structure which can cause localization of phonons.[22] In addition, we observe the E$_{1g}$ peak (MoTe$_2$: 107.1 cm$^{-1}$ and MoSe$_2$: 172.8 cm$^{-1}$) in both our MBE films. While this peak is absent in exfoliated MoTe$_2$, there is a faint signature in exfoliated MoSe$_2$. However, Kan et al.[33] have shown from density-functional calculations that a stable E$_{1g}$ mode does exist around 110 cm$^{-1}$ for 2H-MoTe$_2$. Further, the existence of the B$^1_{2g}$ peaks (MoTe$_2$: 282.4 cm$^{-1}$ and MoSe$_2$: 354.3 cm$^{-1}$) indicates that our films are few-layer thick. The unlabeled peak at around 138 cm$^{-1}$ in MoTe$_2$ could be a second-order Raman mode that is observed even in the exfoliated flake, and in prior reports.[15,32] Figure 3(c) shows the photoluminescence (PL) spectrum of the MoSe$_2$ film. The characteristic peak at 1.55 eV corresponds to the direct band-gap transition of few-layer MoSe$_2$, as reported in literature.[34] The ultraviolet-visible (UV-Vis) absorption spectrum of MoSe$_2$ also shows a band-gap close to 1.58 eV (Supporting Information S4). The corresponding PL peak of MoTe$_2$ (~ 0.9 eV) was beyond the measurement window of our setup.

Transmission Electron Microscopy (TEM) was done to evaluate the micro-structure of the films. Plan-view TEM was employed to study the surface lattice arrangement of atoms, and cross-sectional TEM (X-TEM) for investigation of the film-substrate interface,

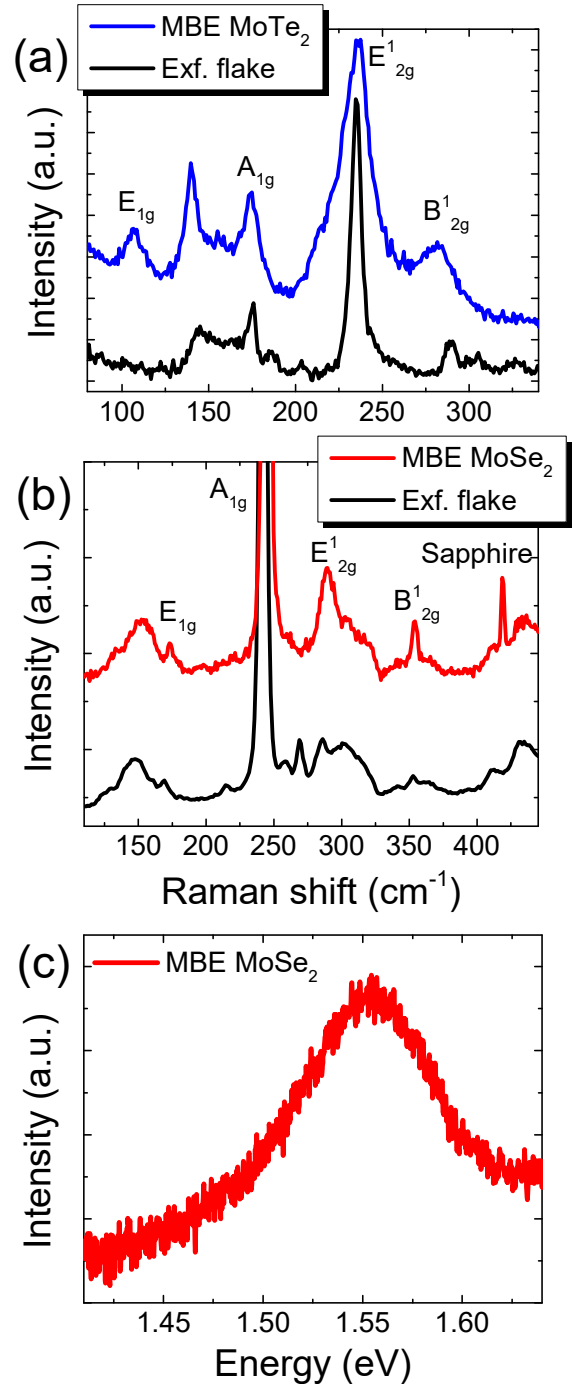

**Figure 3** (a) The Raman spectrum of a 5 nm MoTe$_2$ film (in blue) matches well with the corresponding spectrum of an exfoliated few-layer 2H-MoTe$_2$ flake (in black). The slightly broader peaks in the MBE film could be due to defects. (b) The Raman spectrum of a 5 nm MoSe$_2$ film (in red) and an exfoliated few-layer 2H-MoSe$_2$ flake (in black) also match closely. (c) The PL spectrum of MoSe$_2$ shows a characteristic peak at ~ 1.55 eV.

and layer structure. For plan-view TEM, the films were transferred onto a holey carbon grid using a wet transfer technique. Low-magnification X-TEM images along the <0110> direction show uniform film growth without discontinuities or pinholes, with a thickness around 4-5 nm for both $MoTe_2$ and $MoSe_2$ (Supporting Information S5).

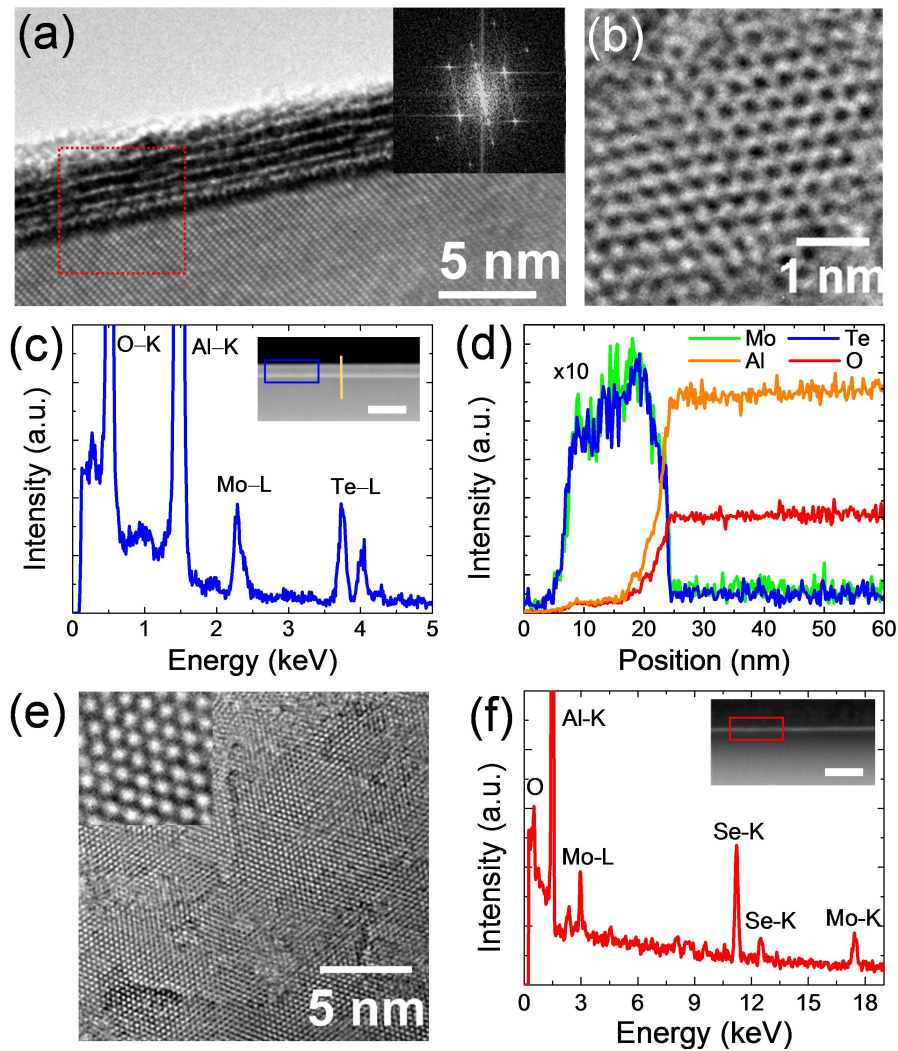

**Figure 4** (a) High-resolution X-TEM image of $MoTe_2$ shows the layer structure confirming vdWE growth. The inset shows the fast Fourier transform of the region marked in red. (b) The plan-view TEM image of $MoTe_2$ shows the hexagonal arrangement of surface atoms. (c) The EDX area spectrum acquired from the STEM image (inset) shows signals from the constituent elements. (d) The EDX line spectrum acquired along the line marked in the inset of (c). (e) Plan-view TEM image of $MoSe_2$ shows its hexagonal lattice (zoomed in view in the inset is $2 \times 2$ nm$^2$). (f) The EDX area spectrum of $MoSe_2$ shows contribution from the region marked in the inset.

Figure 4(a) shows a representative high-resolution X-TEM (HR-XTEM) image for a 4 nm $MoTe_2$ film. The distinct layered structure of $MoTe_2$ is apparent, along with its abrupt interface with the sapphire substrate. The interlayer spacing of ~ 7 Å agrees closely with the bulk $MoTe_2$ spacing.[27] The contrast variation in the layer closest to the substrate could be attributed to strain relaxation in the $MoTe_2$. The presence of grain boundaries and dislocations could be the reason for contrast variations in the layers further from the interface. The inset of Figure 4(a) shows the fast Fourier transform pattern obtained from the area marked in red, which highlights the hexagonal crystal structure and epitaxial relation between the film and the substrate. The hexagonal lattice arrangement of atoms is shown in the plan-view TEM image in Figure 4(b). The lattice constant ($a = 3.5$ Å) extracted from the plan-view TEM micrographs matches with that of bulk 2H-MoTe$_2$.[27] Plan-view TEM images of a larger area show the presence of grains and defects (Supporting Information S6). However, considering the possibility of film damage during transfer onto the TEM grid, it is difficult to precisely comment on the origin of these grains and defects. Similar grain structure has been observed in previous reports of MBE MoSe$_2$ films and been attributed to the low mobility of Mo on $c$-Al$_2$O$_3$(0001) surfaces.[22,35] Figure 4(c) and (d) show the energy dispersive x-ray spectroscopy (EDX) scans of the MoTe$_2$ film from the marked regions of the scanning transmission electron microscopy (STEM) image shown in the inset of Figure 4(c). The EDX spectrum from the area highlighted in blue shows peaks corresponding to Mo, Te (from the MoTe$_2$ film) and Al, O (from the sapphire substrate). Figure 4(d) shows the EDX profile along the red line marked in the inset of Figure 4(c). The signals from Mo and Te show the highest intensity at the location corresponding to the MoTe$_2$ film, and fall abruptly away from the film. Signals from Al and O dominate in the regions away from the MoTe$_2$ film. An atomic percentage ratio of Mo:Te ~ 1:1.86 is extracted from the EDX spectrum, which is consistent with the ratio calculated from *in situ* XPS analysis. Figure 4(e) shows the plan-view TEM image of MoSe$_2$. The lattice arrangement of atoms is hexagonal throughout the image (as shown in a zoomed-in view in the inset), albeit with the existence of defects and grains. The extracted lattice constant of 3.3 Å matches well with the bulk 2H-MoSe$_2$, and the interlayer spacing is ~ 6.7 Å (Supporting Information S6), which agrees well with the data obtained from XRD.[28] Figure 4(f) shows the EDX spectrum of the region marked in the inset, which gives an Mo:Se ratio of ~ 1:1.94.

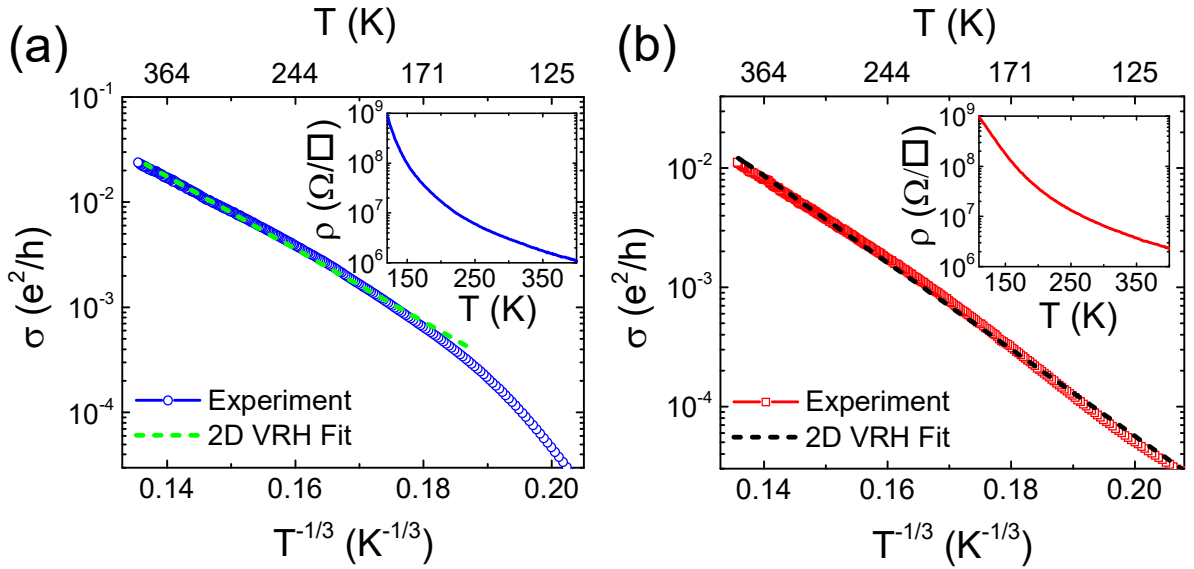

**Figure 5** Electrical transport measurements of (a) MoTe$_2$ and (b) MoSe$_2$. The variation of conductivity on a semi-log scale vs T$^{-1/3}$ follows a linear trend which can be fit to a 2D Mott VRH transport model (dotted lines). The temperature dependence of the measured resistivities (shown in the insets) follow an insulating trend.

We now move on to electrical characterization of the films. The large-area continuous nature of the films, and the insulating sapphire substrate enabled us to perform temperature-dependent resistivity measurements of the as-grown films using a four-point probe geometry. Both the MoTe$_2$ and MoSe$_2$ showed a similar insulating behavior i.e., resistivity increasing with decreasing temperature (insets of Figure 5(a) and (b)). The data was initially fit with an Arrhenius law dependence, $\sigma \propto e^{-E_a/k_BT}$, where $\sigma$ is the conductivity, $E_a$ is the thermal activation energy, $k_B$ is the Boltzmann constant, and $T$ is the temperature. A poor fit, and unreasonably low values of $E_a$ for both the films (Supporting Information S7) indicate that the transport is likely to be dominated by a different physical transport mechanism. We therefore consider a 2D Mott variable-range hopping (VRH) mechanism to describe the transport in our films. Transport mediated by VRH is commonly observed in semiconducting films with a predominance of localized charge-carrier states.[26,36,37] The temperature dependence of $\sigma$ in VRH follows

$$\sigma(T) \propto \exp\left[-\left(\frac{T_0}{T}\right)^{1/3}\right]$$

where, $T_0$ is a fitting parameter, and the exponent 1/3 is characteristic to 2D transport. The 2D VRH model has been used to describe transport in disordered TMDs in several prior literature reports.[15,26,38] Figure 5(a) and (b) show the normalized conductance on a semi-log scale plotted with respect to $T^{-1/3}$, for $MoTe_2$ and $MoSe_2$, respectively. The linear dependence confirms that the conduction in our films is indeed dominated by a VRH mechanism. We note that the fit for $MoTe_2$ is over a smaller temperature range compared to $MoSe_2$. The extracted values of $T_0$ are $4.95 \times 10^5$ K for $MoTe_2$ and $5.88 \times 10^5$ K for $MoSe_2$. The high values of $T_0$ compared to values reported[26,39] for $MoS_2$ indicate a higher degree of carrier localization, possibly stemming from disorder due to chalcogen vacancies, and/or the grainy structure of our films. The VRH behavior in our films is likely due to intrinsic short-range structural disorder.[26,39] The higher values of $T_0$ also imply a smaller localization length ($\xi$), which impedes charge carriers to hop across localized states. This high degree of localization can lead to VRH type transport even at room temperature.[36] The observation of VRH mediated transport in our films concurs with the existence of grains and defects as manifested in the TEM images. Furthermore, the sensitivity of the top few layers to oxygen and moisture, more so in the case of $MoTe_2$, can exacerbate the pre-existing disorder and affect the transport in the films.[40]

**CONCLUSION**

In conclusion, we have carried out the MBE growth of $MoTe_2$ and $MoSe_2$ thin films on insulating $c$-$Al_2O_3(0001)$ substrates. *In situ* characterization using RHEED showed sharp streaky features indicative of atomically flat films grown *via* a vdWE growth mechanism. The film stoichiometry and chemical composition were investigated through *in situ* XPS measurements, and revealed homogeneous phases of $MoTe_2$ and $MoSe_2$, albeit being slightly chalcogen deficient. XRD measurements showed peaks characteristic of the 2H polytypes for both $MoTe_2$ and $MoSe_2$, and further confirmed growth along the $c$-axis of the films. Raman spectroscopy measurements on the grown films match closely with the corresponding bulk flakes, highlighting their crystalline quality. Cross-sectional and plan-view TEM measurements were used to conclusively determine the layer structure and hexagonal arrangement of surface atoms. Finally, electrical measurements on the as-grown films showed an insulating behavior, which was found to follow a 2D Mott variable-range hopping mechanism, suggestive of localized charge carrier states arising from disorder.

The large difference in vapor pressures between Mo and the chalcogen species severely tightens the growth window, in turn translating to a high propensity for chalcogen defects and a diminished grain size.[25] The resulting disorder in the films is apparent from both the TEM images and transport measurements. Our results suggest that despite their seemingly promising spectroscopic quality, MBE-grown TMDs on sapphire may be prone to microscopic defects and disorder, which degrade their electrical characteristics. The choice of substrate, growth conditions, and post-growth treatment play a major role on the film quality, and therefore merit further investigation.

## MATERIALS AND METHODS

**Growth:** The growth was done in a custom-built MBE growth chamber (Omicron, Germany) under ultra-high vacuum (UHV) conditions (base pressure ~$1\times10^{-10}$ mbar). Details of the system have been described elsewhere.[41] After pre-cleaning in acetone and isopropanol, the insulating $c$-$Al_2O_3$(0001) substrates were introduced into the UHV chamber and prepared by resistive heating at 600 °C for 3 hours followed by 700 °C for 30 min, and monitored *in situ* by RHEED. Molybdenum and tellurium (selenium) fluxes generated by an *e*-beam evaporator and an effusion cell (hot lip Knudsen, NTEZ, Scienta Omicron), respectively, were co-deposited onto the substrates at a substrate temperature of 350 °C (250 °C). The chamber pressure during growth never exceeded $1\times10^{-9}$ mbar and the $Te_2$/Mo ($Se_2$/Mo) beam equivalent pressure (BEP) flux ratio was kept around 15. The growth was followed by a post-deposition *in situ* anneal at 600 °C for 10 min with the chalcogen flux ON, which resulted in greatly improved streak features in RHEED. Several samples with thicknesses varying from 3 nm to 10 nm were grown with typical growth rates around 0.1 nm/min.

**Characterization:** Post-growth investigations were carried out by *in situ* RHEED operated at 13 kV, and XPS with a monochromatic Al-K$\alpha$ source ($h\nu$ =1486.7 eV) operating at 15 kV. A Renishaw inVia Raman microscope with a 532 nm diode laser was used for the *ex situ* Raman (3600 *l/mm*) and PL (1200 *l/mm*) measurements. A Philips X-Pert XRD system equipped with a Cu X-ray filament source and a PW-3011/20 proportional detector was used for the XRD measurements. The TEM images were taken using FEI-Tecnai TF30 and TF20 microscopes. The plan-view TEM samples were prepared by transferring the as-grown film onto holey carbon TEM grids using a poly(methyl methacrylate)-based wet-transfer technique, by detaching the films from the sapphire substrate using an NaOH etch.

**Electrical Measurements:** The as-grown samples were contacted at the four corners using indium dots to perform four-point measurements. Temperature dependent measurements were performed in a Lakeshore CPX probe-station using an Agilent B1500A parameter analyzer. The measurements were done in a constant current mode, with a dc excitation current of 5 nA.

*Conflict of interest:* The authors declare no competing financial interest.

*Acknowledgment:* This work was supported in part by NRI SWAN and NSF NNCI. We appreciate technical support from Omicron.

*Supporting Information Available:* S1: Large-area MBE Growth, S2: Effect of Annealing: Evolution of RHEED, S3: X-ray Photoelectron Spectroscopy Survey Scan, S4: UV-Vis Absorption Spectroscopy, S5: X-TEM of $MoTe_2$ and $MoSe_2$, S6: Plan-view TEM of $MoTe_2$ and $MoSe_2$, S7: Arrhenius Fit to Electrical Data.

**Figure TOC**

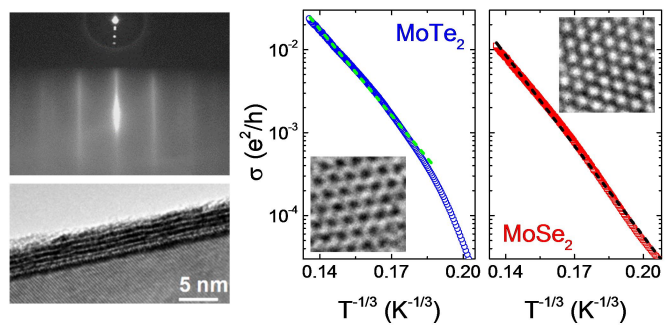

# Supporting Information

**S1: Large-area MBE Growth**

Figure S1 shows optical micrographs of the large-area MBE grown $MoTe_2$ and $MoSe_2$ thin films on sapphire substrates. The thicknesses of the $MoTe_2$ and $MoSe_2$ films are about 5 nm and 7 nm, respectively. The two transparent stripes on the vertical edges correspond to the sample holder clamps, and therefore have no growth.

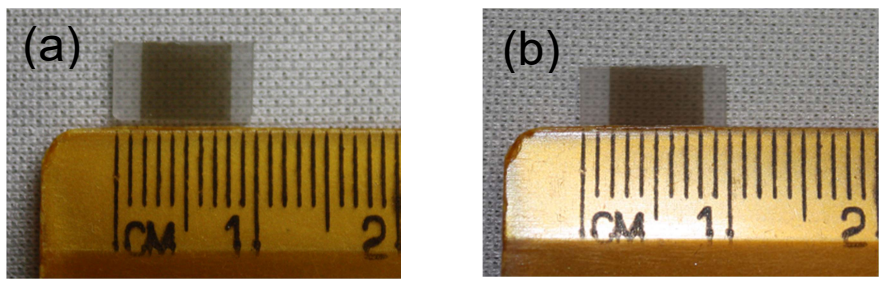

**Figure S1** Optical micrographs of MBE grown (a) $MoTe_2$ and (b) $MoSe_2$ thin films on sapphire.

**S2: Effect of Annealing: Evolution of RHEED**

The as-grown $MoTe_2$ and $MoSe_2$ films showed diffused RHEED features. A post-growth anneal was done to improve the film quality. Figure S2 shows the RHEED images before and after annealing at 600 °C for 10 min in a chalcogen rich environment. There is a marked improvement in the sharpness and streakiness for both $MoTe_2$ and $MoSe_2$, indicating an atomically flat surface morphology. The high temperature anneal increases the surface diffusion of atoms and the presence of a chalcogen flux ensures healing of defects arising from vacancies.[1] The films grow along the *c*-axis of the substrate as should be the case for van der Waal's epitaxy.

**S3: X-ray Photoelectron Spectroscopy Survey Scan**

Figure S3(a) and (b) show the *in situ* XPS survey scans of the $MoTe_2$ and $MoSe_2$ thin films, respectively. All the major peaks have been identified and assigned to Mo, Te (Se). The C-1s peak from the substrate holder was used to calibrate the peak positions.

**S4: UV-Vis Absorption Spectroscopy**

Room temperature ultraviolet-visible (UV-Vis) absorption spectroscopy has been conducted on a 5 nm $MoSe_2$ thin film. The data was taken using a Cary 5000 UV-Vis NIR spectrometer. Figure S4 shows the absorbance spectrum as a function of wavelength. The two absorption peaks at 786 nm and 697 nm correspond to the two exciton peaks of $MoSe_2$, A and B, respectively, originating from the interband exciton transitions at the K-point of the Brillouin zone.[2]

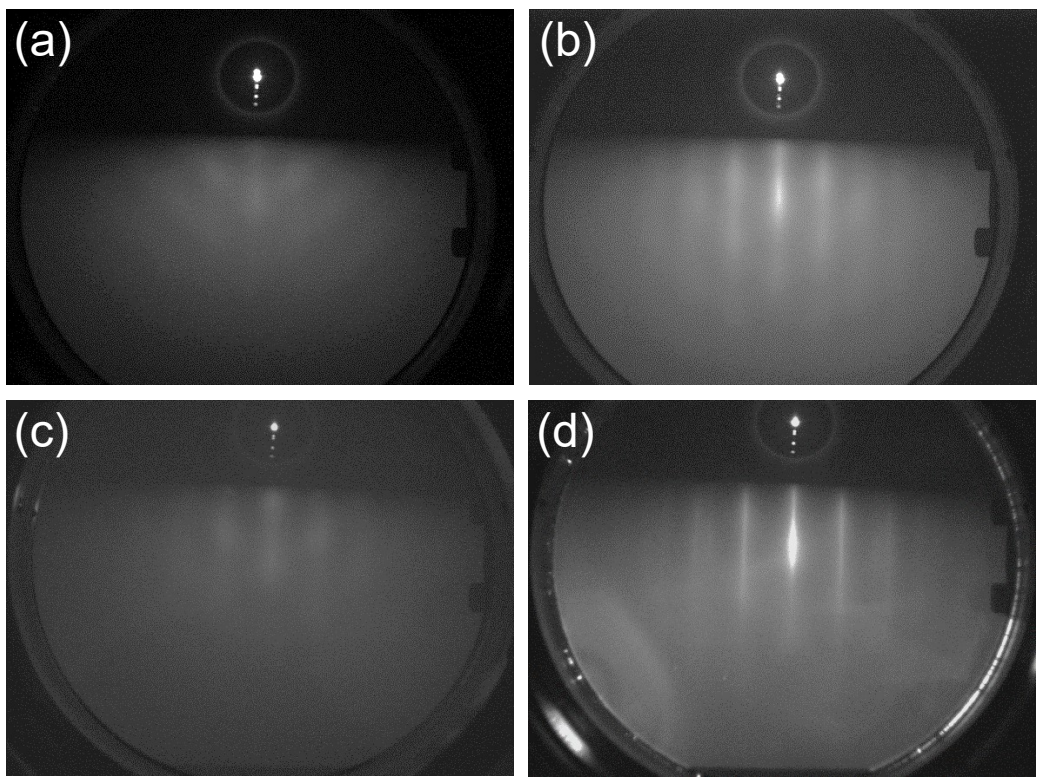

**Figure S2** RHEED patterns along the [1 1 -2 0] direction of sapphire: from as-grown MoTe$_2$ and MoSe$_2$ thin films (a, c) before and (b, d) after annealing, respectively.

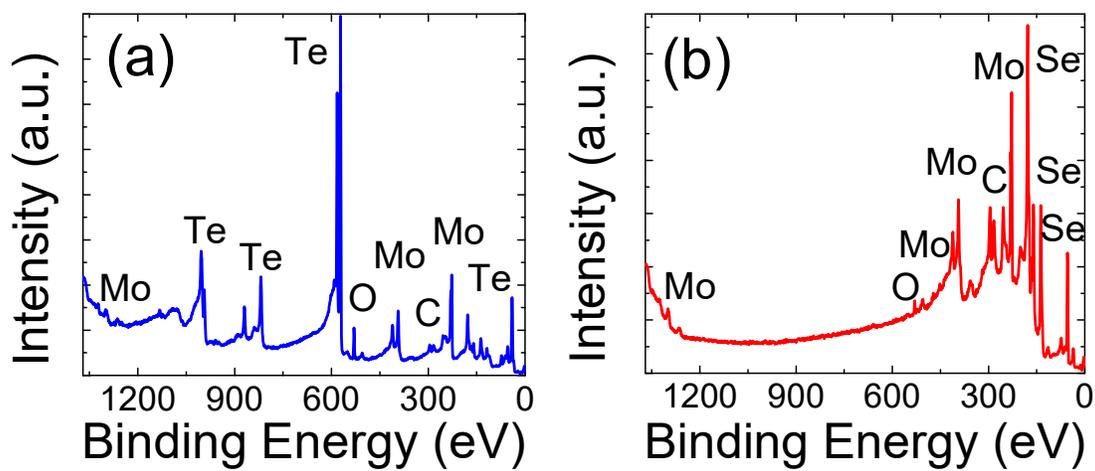

**Figure S3** *In situ* XPS survey scans from (a) MoTe$_2$ and (b) MoSe$_2$ thin films.

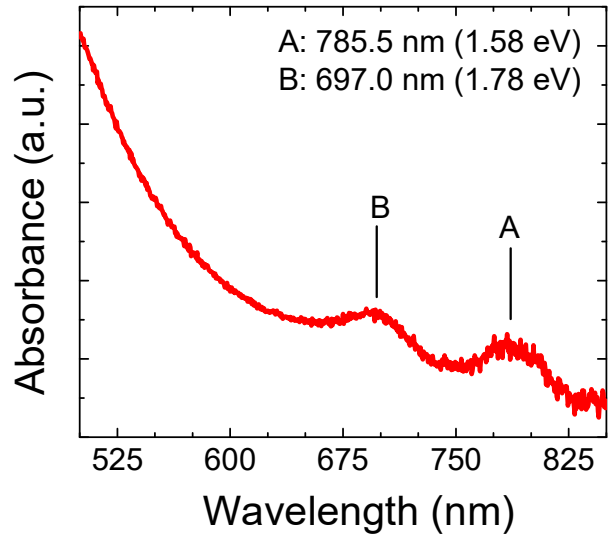

**Figure S4** UV-Vis absorption spectra of a 5 nm MoSe$_2$ thin film at room temperature show the two characteristic exciton peaks, A and B.

### S5: X-TEM of MoTe$_2$ and MoSe$_2$

Figure S5 shows the low-magnification X-TEM micrographs of the MoTe$_2$ and MoSe$_2$ thin films. The thickness of the films used for X-TEM was around 4-5 nm. The films seem to be continuous without any pinholes or discontinuities. The higher magnification X-TEM image in Figure S5(d) shows grain-like features, with grain sizes around 10-20 nm.

### S6: Plan-view TEM of MoTe$_2$ and MoSe$_2$

Figure S6(a) shows the as-captured plan-view TEM image of MoTe$_2$ transferred onto a holey carbon grid. The grain structure is apparent. Due to high reactivity of MoTe$_2$ in ambient conditions, some of the defects are likely to be generated during the transfer process. Figure S6(b) shows the plan-view TEM image of MoSe$_2$ in a region with a wrinkle generated during the transfer process, thereby exposing the layered structure. An interlayer spacing of ~ 6.7 Å is extracted.

### S7: Arrhenius Fit to Electrical Data

Figure S7(a) and (b) show Arrhenius fits to the temperature dependent conductivity of MoTe$_2$ and MoSe$_2$, respectively. The extracted values of activation energy are abnormally low ($E_a$ ~ 18 meV for MoTe$_2$, and ~ 14 meV for MoSe$_2$), and therefore we use a 2D VRH model.

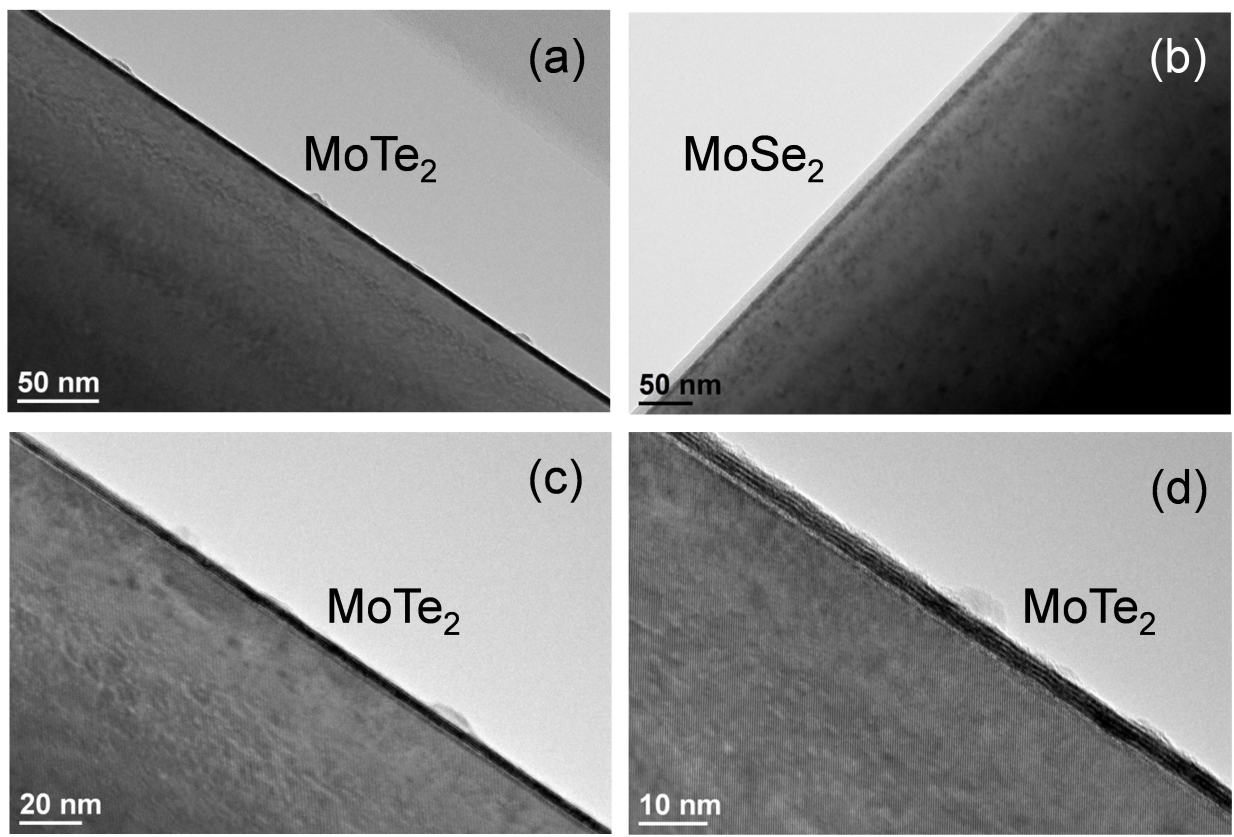

**Figure S5** Low-magnification X-TEM micrographs of MoTe$_2$ and MoSe$_2$ show full film coverage on the substrate. (a) and (b) show large area MoTe$_2$ and MoSe$_2$ films over several hundred nm. (c) and (d) show MoTe$_2$ films at a higher magnification.

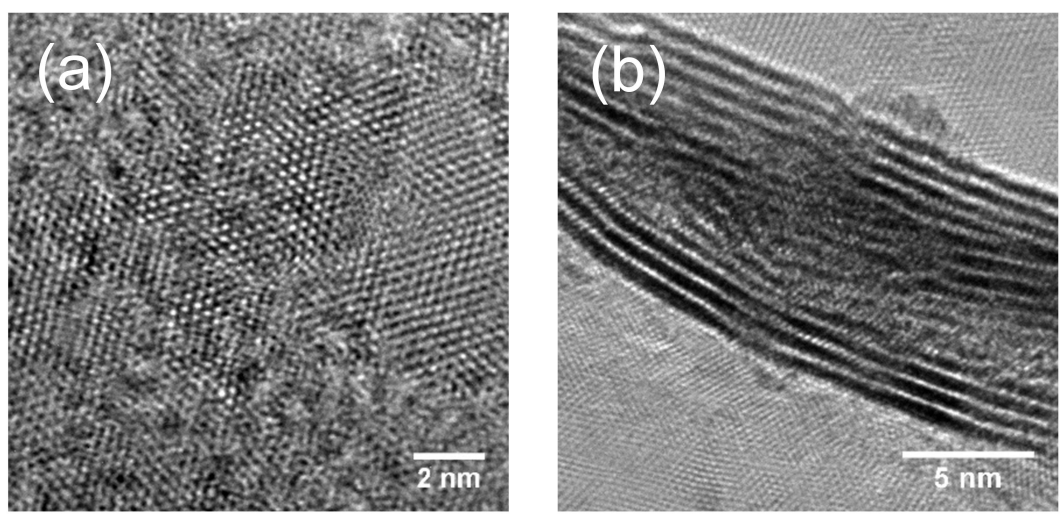

**Figure S6** (a) Plan-view TEM micrograph of MoTe$_2$ shows its grainy structure. (b) Plan-view TEM micrograph of MoSe$_2$ in a wrinkled region uncovers its layer structure.

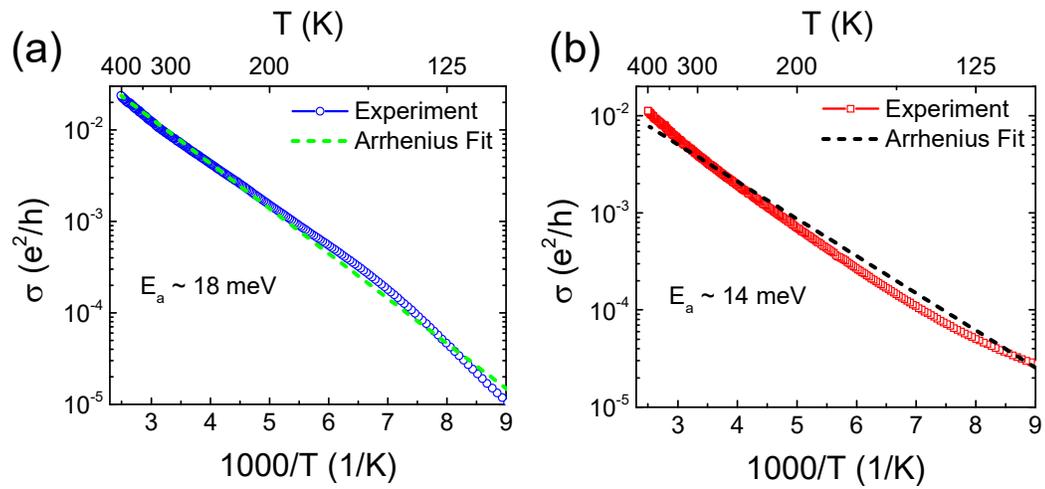

**Figure S7** Arrhenius fits to the temperature dependent conductivity plotted on a semi-log scale for (a) MoTe$_2$ and (b) MoSe$_2$ result in abnormally low values of activation energy.